%% file: vhtau_prl_publish_preprint.tex
\documentclass[aps,preprint,showpacs,groupedaddress]{revtex4}  
\usepackage{graphicx}  
\usepackage{dcolumn}   
\usepackage{bm}        
\usepackage{amssymb}   

\usepackage{xspace}

\def\etal{{\sl et al.}}

\newcommand{\ppb}{\mbox{\ensuremath{p\bar p}}}

\newcommand{\invfb}{fb$^{-1}$}

\newcommand{\pz}{\phantom{0}}

\newcommand{\tautau}{\mbox{\ensuremath{\tau\tau}}}
\newcommand{\taunu}{\mbox{\ensuremath{\tau\nu}}}

\newcommand{\met}{\mbox{\ensuremath{\slash\kern-.7emE_{T}}}}
\newcommand{\mht}{\mbox{\ensuremath{\slash\kern-.7emH_{T}}}}
\newcommand{\mpt}{\mbox{\ensuremath{\slash\kern-.5emT_{T}}}}

\newcommand{\sht}{\mbox{\ensuremath{H_{T}}}}
\newcommand{\pt}{\mbox{\ensuremath{p_{T}}}}

\newcommand{\mh}{\mbox{\ensuremath{M_{H}}}}

\newcommand{\pttau}{\mbox{\ensuremath{p_{T}^{\tau}}}}
\newcommand{\pttrk}{\mbox{\ensuremath{p_{T}^\mathrm {trk}}}}
\newcommand{\ettau}{\mbox{\ensuremath{E_{T}^{\tau}}}}

\newcommand{\mjj}{\mbox{\ensuremath{M_\mathrm{jj}}}}

\newcommand{\ttbar}{\mbox{\ensuremath{t\overline t}}}
\newcommand{\wj}{\mbox{\ensuremath{W+\mathrm{jets}}}}
\newcommand{\zj}{\mbox{\ensuremath{Z+\mathrm{jets}}}}

\begin{document}

\hspace{5.2in} \mbox{Fermilab-Pub-09/089-E}

\title{Search for the standard model Higgs boson in tau final states}

\input{list_of_authors_r2.tex}
\date{March 27, 2009}

\begin{abstract}
We present a 
search for the standard model Higgs boson using hadronically decaying
tau leptons, in 1 fb$^{-1}$ of data
collected with the D0 detector at the Fermilab Tevatron \ppb ~collider.  
We select two final states: $\tau^\pm$ plus missing transverse energy
and $b$ jets, and $\tau^+\tau^-$ plus jets.  These final states are
sensitive to a combination of
associated $W/Z$ boson plus  Higgs boson, vector boson fusion and gluon-gluon
fusion production processes.  The observed ratio of the combined limit on the
Higgs production cross section at the $95\%$ C.L. 
to the standard model expectation is $29$ for a Higgs 
boson mass of $115$~GeV.  
\end{abstract}


\maketitle

A standard model (SM) Higgs boson with a mass in the range $105-145$~GeV
is expected to be produced in \ppb ~collisions at a center-of-mass
energy of 2 TeV 
with cross sections of $\cal O$(100 fb) for 
associated {\it VH} production ($V=W$ or $Z$) 
and vector boson
fusion (VBF), $q\overline q \rightarrow VVq'\overline q''\rightarrow q'
\overline q''H$, 
and of $\cal O$(1 pb) for gluon-gluon fusion (GGF)~\cite{tevhiggs}.  
Previous searches for the SM Higgs boson 
at the Fermilab Tevatron collider~\cite{tevnph} have 
sought the {\it VH} processes with $W/Z$ decays 
to leptons other than taus and $H\rightarrow 
b\overline b$, and the gluon fusion process with $H\rightarrow VV^*$ with 
$V$($V^*$) $\rightarrow ee$ or $\mu\mu$. 
Thus far, there have been no published searches in the case
that either the $V$ or $H$ decays to $\tau$ leptons. 
Given the small Higgs boson production cross sections,
it is advantageous to use all possible decay modes to increase the search
sensitivity.  Here, we present a search
designed for either of the two final states: 
$\tau^\pm \nu + b\overline b$ jets 
(denoted ``\taunu '') or 
$\tau^+ \tau^- +$ jets (denoted ``\tautau ''). 
The analysis is based on 0.94 \invfb ~(\taunu ) and 1.02 \invfb ~(\tautau) 
of data collected by
the D0 experiment~\cite{dzerodet} at the Fermilab Tevatron collider.

The \taunu ~analysis targets 
{\it WH} production with $W\rightarrow \tau\nu$ and
$ZH$ production where $Z\rightarrow \tau\tau$ but 
one $\tau$ is not identified, both with $H\rightarrow b\overline b$. 
The triggers used for selecting events require jets of
high transverse energy, $E_T$, 
and large missing
transverse energy, ~$\met$.  The offline selection of events requires at least 
one tau candidate decaying to hadrons, 
at least two jets identified as candidate $b$ quark jets 
($b$ tagged) with transverse momentum $\pt > 15$~GeV, and $\,\met$, 
corrected for the presence of muons and taus, greater than $30$~GeV. 
We reject events 
containing an electron with $\pt > 15$~GeV or a muon with $\pt > 8$~GeV to
maintain independence from the \tautau ~analysis and 
other SM Higgs boson searches~\cite{tevnph}.

The \tautau ~analysis targets {\it VH} production with 
$Z\rightarrow \tau^+ \tau^-$ and $H\rightarrow b\overline b$ (denoted ``HZ''),
$V\rightarrow q\overline q$ and $H\rightarrow \tau^+ \tau^-$ 
(``WH'' and ``ZH''),  
VBF with $H\rightarrow \tau^+ \tau^-$, and 
GGF with $H\rightarrow \tau^+ \tau^-$ and at least two associated jets.    
We identify one of the taus through its decay 
to $\mu \nu_\tau \overline{\nu}_\mu$ and the other in a hadronic decay mode.
The events satisfy a combination of single muon and muon plus jets
trigger conditions. 
Offline, events are selected~\cite{lq3} 
by requiring exactly one muon 
with $\pt > 12$~GeV,  pseudorapidity $|\eta| < 2.0$, and 
isolated from other tracks and 
calorimeter activity in a cone surrounding the muon track candidate. 
We also require a hadronic tau candidate and at least two jets. 
The $\tau$ and $\mu$ are
required to be of opposite charge for the primary event sample.  Events
containing an electron with $\pt> 12$~GeV are rejected. 

We identify
three types of hadronic taus, motivated by the decays
(1) $\tau^\pm \rightarrow \pi^\pm \nu$, 
(2) $\tau^\pm \rightarrow \pi^\pm \pi^0 ~\nu$, and
(3) $\tau^\pm \rightarrow \pi^\pm \pi^\pm \pi^\mp (\pi^0) \nu$. 
The identifications~\cite{taunn} are based on the number of associated 
tracks  and activity in the electromagnetic (EM) portion of the calorimeter,
both within a cone ${\cal R}=\sqrt{(\Delta\eta)^2+(\Delta\phi)^2}<0.5$,
where $\phi$ is the azimuthal angle. 
The requirements for the \taunu 
~(\tautau ) analysis are:
for type~1, a single track with \pttrk ~$>$ 12 (15) GeV and no nearby EM energy
cluster;   
for type~2, a single  track with \pttrk ~$>$ 10 (15) GeV with an associated 
EM cluster, and for
type~3, at least one track with \pttrk ~$>$ 7 GeV and 
$\Sigma p_{T}^{\rm trk} > 20$~GeV  
and an associated EM cluster.  
In addition to hadronic tau decays, type~2 taus also 
contain $\tau\to e$ decays.
Due to the larger multijet background, type~3 taus are not used in the
\taunu ~analysis.   For the $\tau\tau$ channel only those two-track type~3
candidates with both tracks of the same charge sign are retained to give 
unambiguous tau charge determination. 
A neural network (NN)~\cite{taunn} 
is formed for each tau type using input variables 
such as isolation and the transverse and longitudinal shower profiles
of the calorimeter energy depositions associated with
the tau candidate.  Tau preselection is based on the requirement that
the output NN value, {\sl NN}$_\tau$, 
exceeds $0.3$ thus favoring the tau hypothesis. 
The tau transverse momentum \pttau ~is constructed from 
the transverse energy observed in the calorimeter, \ettau , 
with type-dependent corrections based
on the tracking information.
For the three types we require \pttau ~to 
be greater than 12 (15), 10 (15), or (20)~GeV 
for the \taunu ~(\tautau ) analyses.  
The \taunu ~analysis subdivides the type~2 taus according to whether
the energy deposit is electron-like or hadron-like and the two subsamples
are treated separately in assessing the multijet background.   
For type 2 candidates in the \tautau ~analysis, we require 
$0.7 < \pttrk /\ettau < 2$ to remove backgrounds in regions
with poor EM calorimetry or due to cosmic rays.

Jets are reconstructed with a cone of radius 0.5 in 
rapidity-azimuth space~\cite{jetalgorithm}. 
Their energies are corrected to the particle level 
to account for detector effects and missing energy 
due to semileptonic decays of jet fragmentation products. 
We preselect jets with 
\pt ~$>$ 15 GeV, $|\eta|<2.5$, and separated by
$\cal R$ $> 0.5$
from $\tau$ and $\mu$ candidates.  

Backgrounds other than those from multijet (MJ) production are simulated
using Monte Carlo (MC).  We use 
{\sc alpgen}~\cite{alpgen} for 
\ttbar ~and $V$+jets production; 
{\sc pythia}~\cite{pythia} for {\it WW}, {\it WZ} and {\it ZZ} 
(diboson) production; and 
{\sc comphep}~\cite{comphep} for single top quark production.
The {\sc alpgen} events are passed through {\sc pythia}
for parton showering and hadronization. 
The Higgs boson signal processes
are generated using {\sc pythia} and the 
CTEQ6L1~\cite{cteq} leading order parton distribution functions (PDF)
for \mh ~= 105 -- 145~GeV in 10~GeV steps. 
We normalize the cross sections to the highest available order 
calculations for the signal~\cite{mcfm} and background~\cite{hahn}.
Higgs decays are simulated using {\sc hdecay}~\cite{hdecay} 
and for tau decays 
using {\sc tauola}~\cite{tauola}.  All MC events are passed 
through the standard D0 detector simulation, digitization, and reconstruction 
programs.

Backgrounds due to MJ production, with spurious \met ~or 
misidentified taus are estimated from data samples.  
For the \taunu ~analysis, an enriched multijet
sample is formed  
by selecting taus with $0.3 <$ {\sl NN}$_\tau < 0.7$. 
The contributions from those background processes generated by MC 
are then subtracted to give the BG$_{\tau\nu}$ multijet background 
sample which has
negligible Higgs boson signal and provides 
the shapes of the multijet distributions
in the kinematic variables.  The normalization is given 
by the ratio of the number of events in the signal region,
{\sl NN}$_\tau > 0.9$, after subtracting MC backgrounds, to the number of 
events in the BG$_{\tau\nu}$ sample.  
 
For the MJ background in the \tautau ~analysis, we prepare a multijet
background data sample (BG$_{\tau\tau}$), 
orthogonal to the signal sample (SG$_{\tau\tau}$) defined
by the $\mu$, $\tau$, and jet preselection cuts above,  
by reversing both track and calorimeter 
isolation requirements
for the muon and by requiring {\sl NN}$_\tau < 0.8$. 
For both BG$_{\tau\tau}$ 
and SG$_{\tau\tau}$ samples, 
the MC backgrounds are subtracted, and
the same sign (SS) or opposite sign (OS)  
$\mu$ -- $\tau$ charge combinations subsets are formed. The 
BG$_{\tau\tau}$ sample provides the shape of the multijet background, 
with the normalization obtained by multiplying the number of SS 
SG$_{\tau\tau}$ events by the 
ratio of OS to SS events in the BG$_{\tau\tau}$ sample.  These ratios
are determined separately for each tau type, and are observed to 
be close to one and independent of $p_T^\mu$ and $p_T^\tau$.

The event sample for the \taunu ~analysis is obtained with  
additional requirements after the object selections described above: 
(a) at least two jets with $p_T > 20$~GeV and $\leq 3$ 
  jets with \pt ~$>15$~GeV; 
(b) the angle $\Delta\phi(\met,\mpt) < \pi /2$, where \mpt ~is the 
  negative of the transverse component of the net momentum of 
  all tracks in the event~\cite{mttmet}; 
(c) \sht ~$<200$~GeV, where \sht ~is the scalar sum of the $\pt$ of all jets; 
(d) for hadron-like type 2~taus, the transverse mass, 
  formed from the $\tau$ and $\,\met$, less than $80$~GeV;
(e) dijet invariant mass in the range $50 < M_{\rm jj} < 200$~GeV; and
(f) the requirement  $\Delta\phi(\tau, \met )<0.02(\pi-2)(\met -30) + 
2$ (\met \,in GeV)
 to reduce contamination due to poorly reconstructed multijet events
in which a jet misidentified as a tau is nearly collinear with \met .
To further improve the signal (S) over background (B) separation,
we require two jets to be tagged with a NN that discriminates $b$ quark
jets and jets from light partons~\cite{btag}. 
Figure~1(a,b) shows the \mjj ~distribution before and after $b$ tagging
and the event yields are summarized in Table~\ref{tab:yields}. 

\begin{figure*}[t]
\begin{center}
\includegraphics[width=0.325\textwidth]{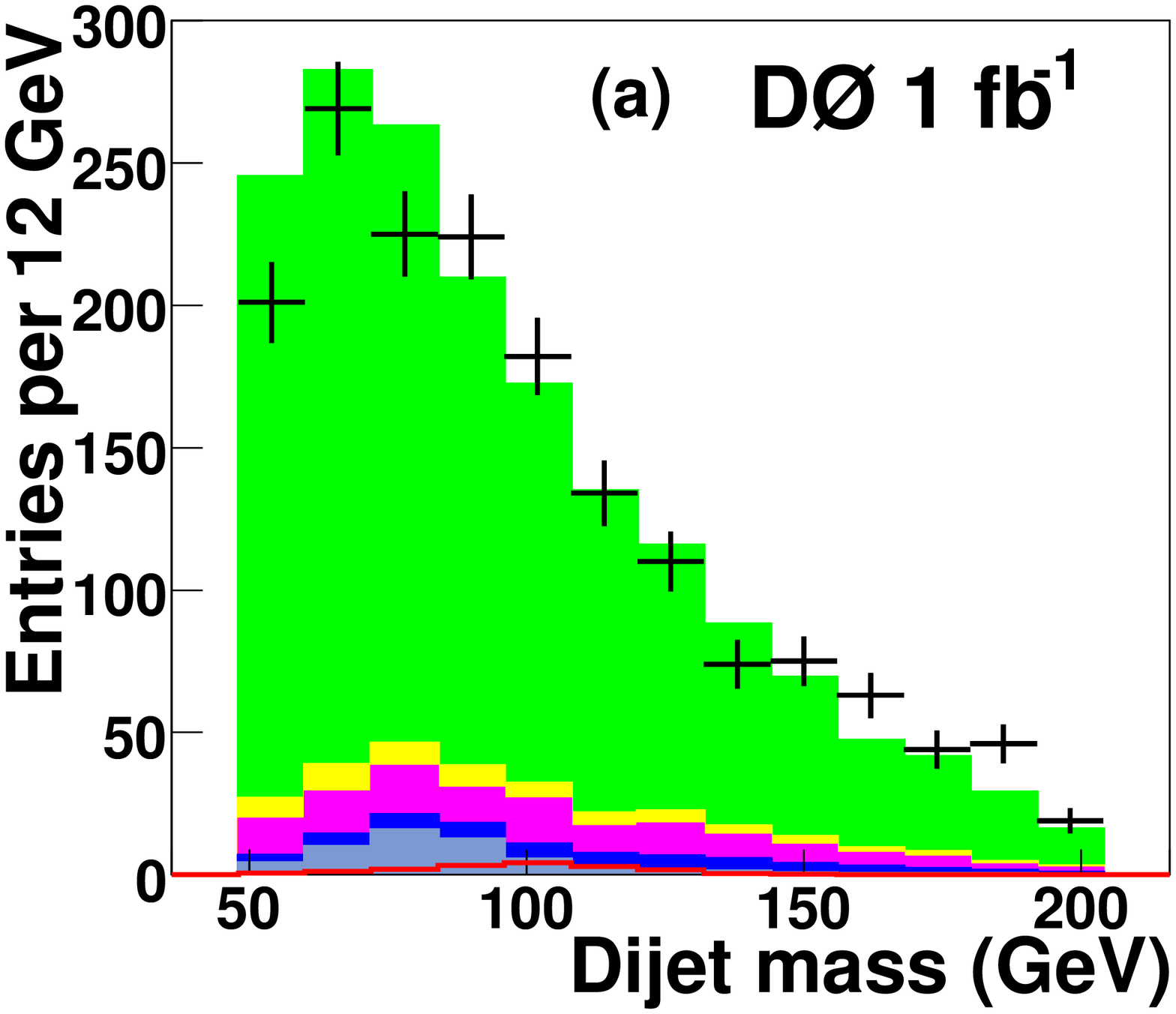}
\includegraphics[width=0.325\textwidth]{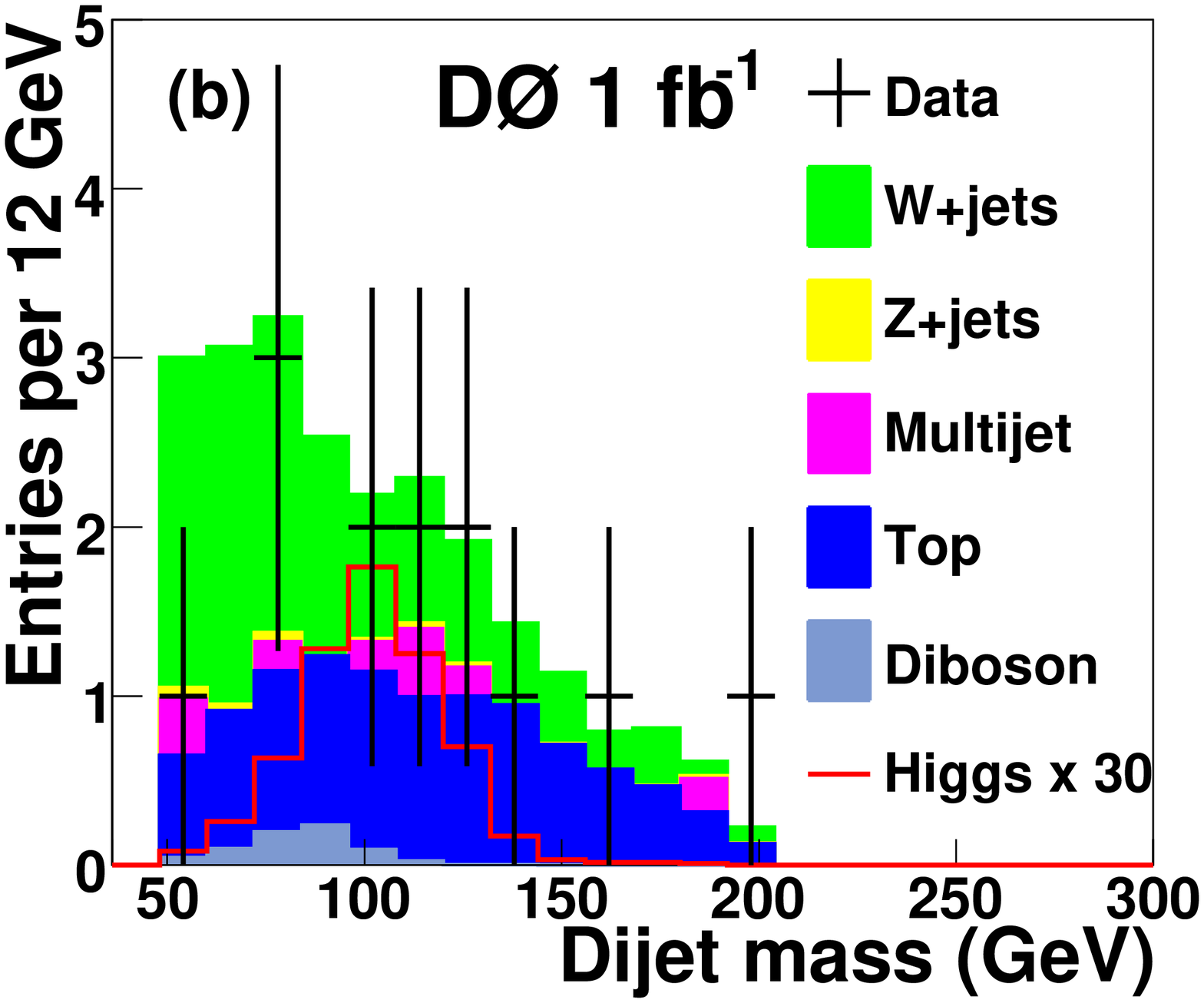}
\includegraphics[width=0.325\textwidth]{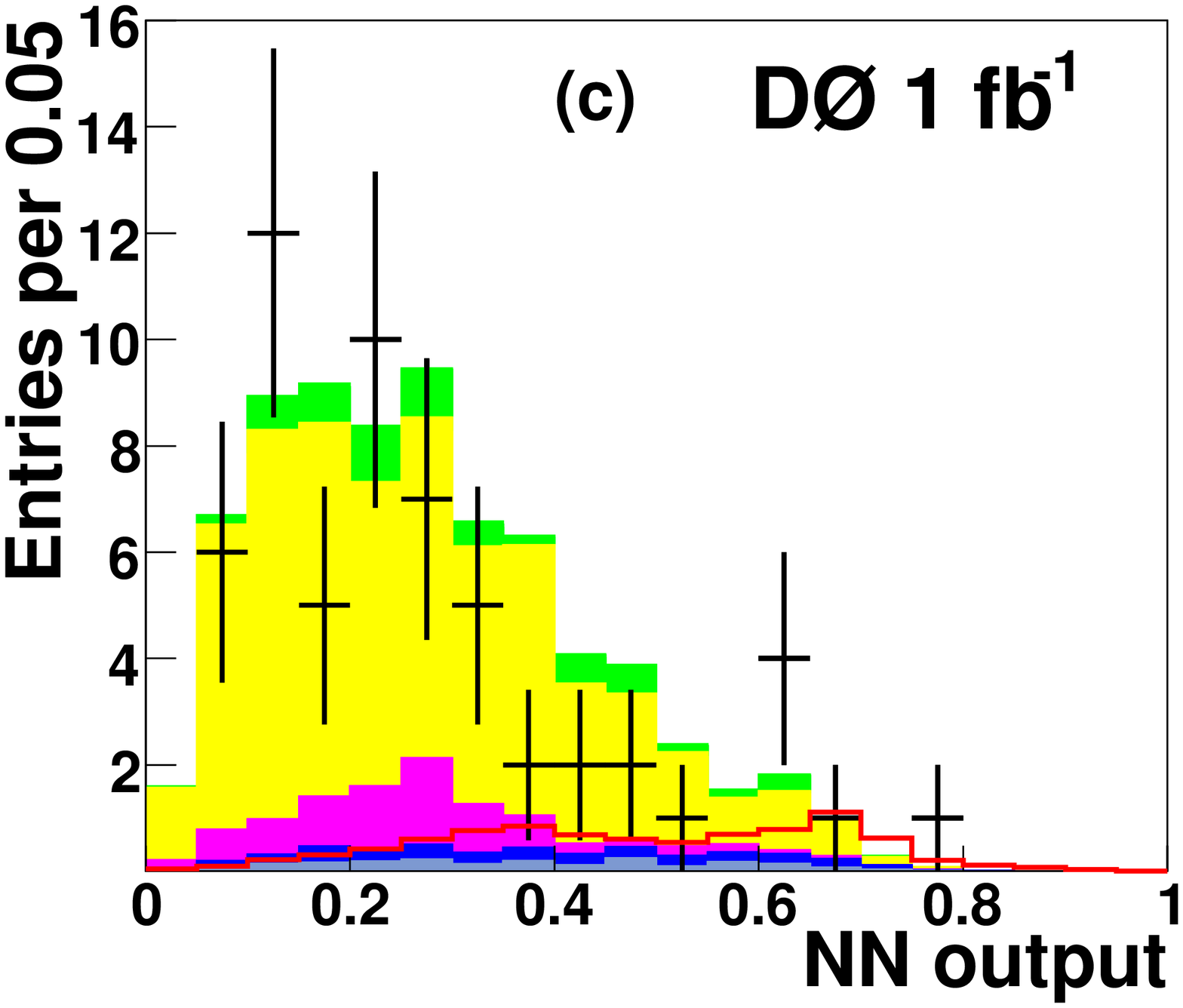}
\caption{\label{fig:fig1} 
The dijet mass distribution for all tau types for the 
\taunu ~analysis (a) before $b$-tagging, and (b) after the final selection;
(c) the combined {\sl NN}$_{\rm Zjets}$ ~variable 
for the low Higgs mass \tautau ~analysis. The signal is shown (multiplied by
30) for $M_H=115$ GeV.  (color online)
} 
\end{center}
\end{figure*}

Most of the signal processes sought in the \tautau ~analysis contain
light quark jets, so we do not
employ $b$ tagging.
We require 2 jets with $p_T>20$ GeV.
To further separate signals from backgrounds, 
we train a dedicated NN for the signal processes 
(HZ, WH, ZH, VBF) and for each of the main
background types (\wj , \zj , \ttbar ~and MJ).
After requiring two jets, the MC GGF samples are small, making NN training
unreliable.  Since the GGF and VBF processes both involve non-resonant
dijet systems, we incorporate the GGF events with the VBF sample when
constructing the final limit analysis. 
The NNs are separately trained for low 
mass (105, 115 and 125 GeV) and high mass (135, 145 GeV) Higgs bosons, 
giving 32 NNs in all.  Twenty well-modeled input variables 
are considered for each of the 
NNs.  They include transverse or invariant masses of 
combinations of jets and leptons, \met , 
angular correlations, and overall event distributions such as 
\sht ~and aplanarity\cite{aplanarity}.  
For each signal-background pair, a choice of six or
seven variables is made using the criterion that each added variable must
give significant improvement in $S/\sqrt{B}$.
The same variable
choices are made for all Higgs boson masses.
All NN input and output variables show good agreement between data
and background prediction, and typically provide good discrimination between
the signal and background under  consideration.
The $t\overline t$, $W$+jets and MJ {\sl NN}s give good separation of 
signal and background, whereas the $Z$+jets {\sl NN} signal
and background distributions are not so well differentiated.
Thus we define the variables {\sl NN}$_{\rm bg}$ as the largest NN output
variable among the various signals, for each background source, 
bg = \ttbar , \wj , and MJ. 
We require {\sl NN}$_{\rm bg} > 0.4$, 
based on an optimization of the expected Higgs boson
cross section limits.  
After this selection, the  
NN outputs trained against the \zj ~background 
for all signals are combined by taking their weighted average,
{\sl NN}$_{\rm Zjets}$,
over the four signal processes (HZ, WH, ZH, VBF), with weights equal to
the relative expected yield for each signal. 
 The {\sl NN}$_{\rm Zjets}$ distribution for the final sample is shown 
in Fig.~1(c), now including  the GGF signal events.  
The signal and background event yields are given in Table~\ref{tab:yields}.  

\begin{table}[bt]
\caption{\label{tab:yields}
Numbers of events at the preselection level and 
after the final selection ($b$ tagging for $\tau\nu$ 
and {\sl NN}$_{\rm bg}$ cut for $\tau\tau$)
for all  $\tau$ types combined, 
for data, 
estimated backgrounds and signal at \mh ~= 115 GeV. 
The $V$+jets background is given for light parton (``u,d,s,g'' = ``lp'') and
heavy flavor (``b,c'' = ``hf'') jets separately.
The uncertainties shown are statistical only. For the \taunu ~(\tautau ) 
analysis the combined statistical and systematic uncertainties
on the sum of backgrounds in 
the final selections are 5.3 (14.8) events.}
\begin{tabular}{l|cc|cc} 
\hline
\hline
 & \multicolumn{2}{c}{$\tau\nu$ analysis}
  & \multicolumn{2}{c}{$\tau\tau$ analysis} \\
 \cline{2-5}
   Source & Preselection & Final & Preselection & Final \\ \hline
$W$+ lp  &  $1124\pm18$     & $ \pz 0.5\pm0.0 $   & $\pz 37.7\pm2.1$  & $ \pz 5.1\pm0.3 $   \\
$W$+ hf  &  $ 308.2\pm4.8 $   &  $ 10.9\pm0.3 $ & $\pz\pz 8.2\pm0.5 $  &  $ \pz 0.9\pm0.1 $ \\
$Z$+  lp  &  $\pz 49.1\pm1.5 $     &  $ <0.2 $ & $\pz 78.4\pm0.9 $  &  $ 43.8\pm0.6 $ \\
$Z$+  hf  & $ \pz\pz 7.8\pm0.5 $   &  $\pz  0.4\pm0.0 $ &   $\pz 15.7\pm1.0 $  &  $ 10.1\pm0.7 $ \\
\ttbar        &  $\pz 46.7\pm0.4$      & $\pz 9.5\pm0.1$    &  $\pz 30.8\pm0.3$ & $ 2.8\pm0.0$  \\
Diboson  &  $ \pz 54.9\pm1.1 $    &  $\pz 0.7\pm0.0 $  & $\pz\pz 6.1\pm0.5 $  &  $\pz 2.1\pm0.2 $ \\
Multijet    &  $ \pz 122.6\pm11.2 $   &  $\pz 1.3\pm0.1 $ & $\pz 57.2\pm8.1 $  &  $ \pz 6.5\pm2.8 $ \\ \hline
Sum         &  $ 1714 \pm22$  &  $23.3\pm0.4 $   & $234\pm 9 $  &  $ 71.2\pm3.0 $ \\ \hline
Data  &    $1666$          & $13$           & 220  & 58 \\ \hline
HZ  & & & 0.038 & 0.029\\
WH  & $0.543$     & $0.201$    & 0.145  & 0.106 \\
ZH  & $0.023$    & $0.015$    & 0.094  & 0.069 \\
VBF  & & & 0.071  & 0.059 \\
GGF   & & & 0.041  & 0.030 \\
\hline
Sum & 0.566 & 0.216 & 0.389  & 0.293 \\
\hline \hline
\end{tabular}
\end{table}

Some systematic uncertainties induce a shape 
dependence on the final limit setting variable. 
For the \taunu ~analysis, 
such shape dependence is found for the jet energy scale, 
jet energy resolution, and the $b$-tagging efficiencies. 
 Alternate shapes are
determined by changing the relevant parameter by $\pm 1$ 
standard deviation from the 
nominal value and are provided to the limit setting program.
For the \tautau ~analysis, only the multijet background is found to give 
an appreciable shape change.  It is determined by varying the method for
selecting MJ events, reversing either the muon or the tau requirements, 
but not both, relative to the standard choice.
The remaining `flat' systematic uncertainties do not affect the final variable 
distribution shape.  Such flat uncertainties for the \taunu ~(\tautau )
analysis are, unless otherwise noted, fully correlated for different
backgrounds and analysis channels, and include 
(a) integrated luminosity, 6.1\% (6.1\%)~\cite{lumi}; 
(b) trigger efficiency, 5.5\% (3\%) (uncorrelated \taunu ~and \tautau );
(c) muon identification, (4.5\%);
(d) tau identification, 5.0--6.0\% (5.0\%);
(e) tau track efficiency, 3.0\% (3.0\%);
(f) tau energy scale, 2.3--2.7\% (3.5\%);
(g) jet identification and reconstruction, 1.7--4.9\% (2\%);
(h) jet energy resolution, (4.5\%);
(i) jet energy scale (7.5\%)~\cite{jes}; 
(j) MC background cross sections, 6--18\% (6--18\%) (these are taken 
to be uncorrelated among the backgrounds); 
(k) higher order correction for the $V+$jets cross section, 20\% (20\%);
(l) $V$+ heavy flavor jet cross section correction, 30\% (30\%); 
and
(m) multijet background, 82--100\% (uncorrelated \taunu ~and \tautau ).

The upper limits on the Higgs boson cross section are obtained using
the modified frequentist method~\cite{cls}. 
For the \taunu ~analysis, the test statistic 
is the negative log 
likelihood ratio (LLR) derived from 
the \mjj ~distribution.  For 
the \tautau ~analysis, the LLR is formed from the 
{\sl NN}$_{\rm Zjets}$ final neural network variable.  
The confidence levels {\sl CL}$_{s+b}$
({\sl CL}$_{b}$) give the probability that the 
LLR value from a set of 
simulated pseudo-experiments under the signal plus 
background (background-only) hypothesis is less likely than
that observed, at the quoted C.L.  The hypothesized signal 
cross sections are scaled up from their SM values until the value of 
{\sl CL}$_{s} =$ {\sl CL}$_{s+b}$/{\sl CL}$_{b}$ reaches 0.05 to obtain
the limit cross sections at the 95\% C.L., both for expected 
and observed limits.  
In the calculation, all contributions
to the systematic uncertainty are varied, subject to the constraints given
by their estimated values, to give the best fit~\cite{profiling}. 
The correlations of each systematic
uncertainty among signal and/or background processes are accounted for
in the minimization.

The ratios of the expected and observed upper limits 
to the SM expectations are shown in 
Table~\ref{tab:limits} for the two channels separately and combined.
For all Higgs masses, the observed limits are within 1$\sigma$ of the expected
limits.
At \mh ~= 115 GeV, the observed (expected) 95\% C.L. limit is 
29 (28) times that  predicted in 
the SM for the seven signal processes considered in the
combined \taunu ~and \tautau ~analyses. This is the first
limit on SM Higgs production using final states involving hadronically 
decaying tau leptons.  These results contribute to the sensitivity of the 
combined Tevatron search for low mass Higgs bosons~\cite{tevnph}.

\begin{table}[bt]
\caption{\label{tab:limits}
Expected and observed $95\%$ C.L. upper 
limits on the Higgs boson production cross section
relative to the SM predicted value, for the \taunu ~and \tautau ~analyses
separately and combined.
}
\begin{tabular}{c|cc|cc|cc} 
\hline
\hline
~  & \multicolumn{2}{c|}{\taunu ~ analysis} & \multicolumn{2}{c|}{\tautau ~ analysis} & 
\multicolumn{2}{c}{Combined} \\ \hline
\mh ~(GeV) & exp. & obs. &  exp. & obs. &  exp. & obs. \\ \hline
105 & 33 & 27 & 39 & 36 & 24 & 20 \\
115 & 42 & 35 & 43 & 47 & 28 & 29 \\
125 & 62 & 60 & 60 & 65 & 40 & 44 \\
135 & 105 & 106 & 87 & 61 & 63 & 50 \\
145 & 226 & 211 & 158 & 95 & 120 & 82 \\ 
\hline \hline
\end{tabular}
\end{table}

\input{acknowledgement_paragraph_r2.tex}
\end{document}

%% file: list_of_authors_r2.tex
%
\author{V.M.~Abazov$^{37}$}
\author{B.~Abbott$^{75}$}
\author{M.~Abolins$^{65}$}
\author{B.S.~Acharya$^{30}$}
\author{M.~Adams$^{51}$}
\author{T.~Adams$^{49}$}
\author{E.~Aguilo$^{6}$}
\author{M.~Ahsan$^{59}$}
\author{G.D.~Alexeev$^{37}$}
\author{G.~Alkhazov$^{41}$}
% ...

\affiliation{\vspace{0.1 in}(The D\O\ Collaboration)\vspace{0.1 in}}
% ...
\affiliation{$^{6}$University of Alberta, Edmonton, Alberta, Canada;
                Simon Fraser University, Burnaby, British Columbia, Canada;
                York University, Toronto, Ontario, Canada and
                McGill University, Montreal, Quebec, Canada}
% ...
\affiliation{$^{30}$Tata Institute of Fundamental Research, Mumbai, India}
% ...
\affiliation{$^{37}$Joint Institute for Nuclear Research, Dubna, Russia}
% ...
\affiliation{$^{41}$Petersburg Nuclear Physics Institute,
                St. Petersburg, Russia}
% ...
\affiliation{$^{49}$Florida State University, Tallahassee, Florida 32306, USA}
% ...
\affiliation{$^{51}$University of Illinois at Chicago,
                Chicago, Illinois 60607, USA}
% ...
\affiliation{$^{59}$Kansas State University, Manhattan, Kansas 66506, USA}
% ...
\affiliation{$^{65}$Michigan State University,
                East Lansing, Michigan 48824, USA}
% ...
\affiliation{$^{75}$University of Oklahoma, Norman, Oklahoma 73019, USA}
% ...

%% file: acknowledgement_paragraph_r2.tex
%
We thank the staffs at Fermilab and collaborating institutions, 
and acknowledge support from the 
DOE and NSF (USA);
CEA and CNRS/IN2P3 (France);
FASI, Rosatom and RFBR (Russia);
CNPq, FAPERJ, FAPESP and FUNDUNESP (Brazil);
DAE and DST (India);
Colciencias (Colombia);
CONACyT (Mexico);
KRF and KOSEF (Korea);
CONICET and UBACyT (Argentina);
FOM (The Netherlands);
STFC and the Royal Society (United Kingdom);
MSMT and GACR (Czech Republic);
CRC Program, CFI, NSERC and WestGrid Project (Canada);
BMBF and DFG (Germany);
SFI (Ireland);
The Swedish Research Council (Sweden);
CAS and CNSF (China);
and the
Alexander von Humboldt Foundation (Germany).

%% file: vhtau_prl_publish_preprint.bbl
\begin{thebibliography} {99}
\bibitem{mttmet}
In events with true \met ~due to non-interacting particles, \met ~and \mpt 
~tend to be aligned, whereas for events with mismeasured jets in the 
calorimeter they do not.
\end{thebibliography}
